\documentclass{aa}

\usepackage{graphicx,epsfig}

\def\lsim{\,\lower2truept\hbox{${< \atop\hbox{\raise4truept\hbox{$\sim$}}}$}\,}
\def\gsim{\,\lower2truept\hbox{${> \atop\hbox{\raise4truept\hbox{$\sim$}}}$}\,}
\newcommand{\etal}{{et~al.}}
\newcommand{\ie}{{i.e.}}

\begin{document}


\title{Power spectrum of the polarized diffuse Galactic radio emission}

\author{C.~Baccigalupi\inst{1} \and  C.~Burigana\inst{2} \and
F.~Perrotta\inst{1} \and G.~De~Zotti\inst{3} \and  L.~La~Porta\inst{2} \and
D.~Maino\inst{4} \and M.~Maris\inst{4} \and R.~Paladini\inst{1}}

\institute{SISSA/ISAS, Via Beirut 4, I-34014 Trieste, Italy \and
ITeSRE/CNR, Via P. Gobetti 101, I-40129 Bologna, Italy, \and
Osservatorio Astronomico di Padova, Vicolo dell'Osservatorio
5, I-35122 Padova, Italy \and Osservatorio Astronomico di Trieste,
Via Tiepolo 11, I-34131 Trieste, Italy }

\offprints{C. Baccigalupi (bacci@sissa.it)}

\date{Received / Accepted}


	\abstract{
	We have analyzed the available polarization surveys of the Galactic
	emission to estimate to what extent it may be a serious hindrance to
	forthcoming experiments aimed at detecting the polarized component 
	of Cosmic Microwave Background (CMB) anisotropies. Regions were 
	identified for which independent data consistently indicate that 
	Faraday depolarization may be small.
	The power spectrum of the polarized emission, in terms of antenna
	temperature, was found to be described by
	$C^{P}_{\ell}\simeq (1.2\pm 0.8)\cdot 10^{-9}\cdot
	(\ell / 450)^{-1.8\pm 0.3}\cdot
	(\nu/ 2.4{\rm GHz})^{-5.8}$ K$^{2}$, from arcminute to degree scales.
	Data on larger angular scales ($\ell\le 100$) indicate a steeper 
	slope $\sim \ell^{-3}$.	We conclude that polarized Galactic emission 
	is unlikely to be a serious limitation to CMB polarization measurements 
	at the highest frequencies of the MAP and {\sc Planck}-LFI instruments, 
	at least for $\ell\ge 50$ and standard cosmological models.
	The weak correlation between polarization and total power and the
	low polarization degree of radio emission close to the Galactic plane 
	is interpreted as due to large contributions to the observed 
	intensity from unpolarized sources, primarily strong H{\sc ii} regions, 
	concentrated on the Galactic plane. Thus estimates of the power 
	spectrum of total intensity at low Galactic latitudes are not 
	representative of the spatial distribution of Galactic emission 
	far from the plane. Both total power and polarized emissions show 
	highly significant deviations from a Gaussian distribution.
	}

	\maketitle

\bibliographystyle{astron}

\section{Introduction}
\label{introduction}

Several ongoing or planned experiments (see Staggs \etal 1999 for a
recent review) are designed to reach the sensitivities required to 
measure the expected linear polarization of the Cosmic Microwave 
Background (CMB).

The forthcoming space missions {\sc Planck} and MAP,
aimed at obtaining full sky high sensitivity and high
resolution maps (FWHM of about $56'$, $41'$, $28'$, $21'$,
and $13'$ for MAP channels at 22, 30, 40, 60, and 90 GHz, respectively;
of $33'$, $23'$, $14'$, $10'$ for
{\sc Planck} ``radiometric'' channels at 30, 44, 70, and 100 GHz;
of $10'.7$, $8'$, $5'.5$, $5'$, $5'$, $5'$ for {\sc Planck}
``bolometric'' channels
at 100, 143, 217, 353, 545, and 857 GHz, respectively)
of CMB anisotropies will also probe the CMB
polarization fluctuations (Mandolesi \etal~1998; Puget \etal~1998; 
MAP webpage: {\tt http://map.gsfc.nasa.gov/}; 
{\sc Planck} webpage: 
{\tt http://astro.estec.esa.nl/SA-general/Projects/}-\\
{\tt Planck/}).  
We recall that the CMB radiation brightness peaks at about 160 GHz.

The current design of instruments for the {\sc Planck} mission
(the third Medium-size mission of ESA's Horizon 2000 Scientific
Programme)
provides good sensitivity to polarization at all LFI (Low Frequency
Instrument) frequencies (30--100 GHz) as well as at three
HFI (High Frequency Instrument) frequencies (143, 217 and 545 GHz).
The NASA's MIDEX class mission MAP has also polarization sensitivity
in all channels.

While there is a very strong scientific case for CMB polarization
measurements (cf., e.g., Zaldarriaga 1998 and references therein), they
are very challenging both because of the weakness of the signal and
because of the contamination by foregrounds that may be more polarized
than the CMB.

Our knowledge of polarized foreground components is very meager
(see Davies \& Wilkinson 1999 for a recent review). In this paper
we present a preliminary investigation of the power spectrum
of the polarized Galactic synchrotron emission, the likely dominant
foreground contribution at microwave frequencies where CMB is dominating,
at least at intermediate to large angular scales (Tegmark \etal~2000).
When this work was approaching completion we learned
that a similar analysis was carried out by Tucci \etal~(2000). We improve
on their results by taking into account, in addition to the Parkes survey
(Duncan \etal~1995, 1997; hereafter D97) discussed by them, the more
recent Effelsberg surveys at 2.7 GHz (Duncan \etal~1999; D99) and at 1.4
GHz, covering areas up to $\pm 20^\circ$ of Galactic
latitude (Uyaniker \etal~1998, 1999; U99), as well as the Leiden surveys
(Brouw \& Spoelstra 1976; BS76). We also discuss in some detail the effect of
the emission from H{\sc ii} regions in the Galactic plane and of the Faraday
depolarization.

The Galactic polarized thermal dust emission has been modelled by Prunet \etal~
(1998); based on their results, we may expect that
fluctuations of polarized dust emission prevail over those of
polarized synchrotron emission above 100--150 GHz.
De Zotti \etal~(1999) discussed polarization fluctuations
due to extragalactic sources. Additional polarized contributions are
expected from magneto-dipole emission or rotational emission
of dust grains (Draine \& Lazarian 1999; Lazarian \& Draine
2000) and scattered free-free emission (Keating \etal~1998).
A multifrequency Wiener filtering method to detect CMB polarization in the
presence of polarized foregrounds has been worked out by Bouchet \etal~(1999).

\section{Polarization surveys}
\label{database}

Linear polarization observations extending up to high Galactic latitudes
and carried out at several frequencies, from 408 to 1411 MHz, have been
presented by BS76 (Leiden surveys).
The half power beamwidths varied with increasing frequency from 
$2.3^{\circ}$ to $0.6^{\circ}$.
Unfortunately these and the other surveys discussed by Spoelstra (1984)
are undersampled so that proper smoothing to the largest beamwidth,
as required to combine data at different frequencies, is not possible.
Thus, estimates of differential polarization and differential Faraday
rotation across the beam cannot be made.
We have limited our analysis of these data to patches of the sky
with better than average sampling.

High resolution polarimetric surveys of strips around the
Galactic plane at 2.4 and 2.695 GHz, respectively, have recently been
published by D97 and D99. U99 carried
out a surface brightness and polarization survey of four fields at medium
Galactic latitudes (up to $b=20^\circ$), at 1.4 GHz.
These data, together with details about the instrumental capabilities,
can be found on the WEB site
{\tt http://www.mpifr-bonn.mpg.de/survey.html}.

D97 covered $127^\circ$ of Galactic longitude
($238^\circ < {\rm l} < 5^\circ$) out to at least $b = \pm 5^\circ$
(for $340^\circ < {\rm l} < 5^\circ$ the survey extends to $b = +7^\circ$,
for $340^\circ < {\rm l} < 352^\circ$ to $b = -7^\circ$,
and for $240^\circ < {\rm l} < 270^\circ$ to $b = -8^\circ$) with an angular
resolution of $10.4'$.
The surveyed area amounts to $\simeq 1413$ square degrees.
The nominal rms noise in total power is 17 mJy/beam area
(8 mK), and 11 mJy/beam area (5.3 mK) for polarization; however
a lower rms noise of 11 mJy/beam area (5.3 mK) for total intensity
and 6 mJy/beam area (2.9 mK) for polarization has been achieved
over $\simeq 43\%$ of the total area.
The center frequency is 2.417 GHz and the bandwidth about 145 MHz.
Values of the Stokes $Q$ and $U$ parameters and of the total power
are given every $4'$ on a rectangular grid,
in units of mJy/beam area;
the conversion factor to brightness temperature is
$1\,\hbox{mJy/beam area}= 0.48\,$mK (Duncan, private communication).

Polarimetric data from the Effelsberg 2.695 GHz survey with half-power
beamwidth of $4.3'$ in the first Galactic
quadrant were reported by D99. Maps at a resolution of
$5.1'$ of the Stokes $Q$ and $U$ emission components, covering the region
$74^\circ \geq {\rm l} \geq 4^\circ.9$, $|b|\leq 5^\circ$, with a rms noise
of 2.5 mJy/beam area (corresponding to 9 mK) were constructed.

Of particular interest for our purposes is the continuum and polarization
survey at 1.4 GHz, also carried out with the Effelsberg 100-m
telescope, at medium Galactic latitudes (up to $b=\pm 20^\circ$).
Four areas were observed (with the one in the Cygnus region split in two parts),
totaling about 1050 sq. deg.:
one in the first Galactic quadrant ($45^\circ \geq {\rm l} \geq 55^\circ$,
$4^\circ \geq b \geq 20^\circ$); the northern ($65^\circ \geq {\rm l} \geq
95^\circ$, $5^\circ \geq b \geq 15^\circ$) and the southern ($70^\circ
\geq {\rm l} \geq 100^\circ$, $-15^\circ \geq b \geq -5^\circ$)
parts of the Cygnus region; the highly polarized ``fan region''
($140^\circ \geq {\rm l} \geq 153^\circ$, $3.5^\circ \geq b \geq 10^\circ$);
the anticentre region ($190^\circ \geq {\rm l}
\geq 210^\circ$, $3.8^\circ \geq b \geq 15^\circ$). The rms noise is
about 15 mK (about 7 mJy/beam area) for total intensity and about
8 mK in linear polarization; the angular resolution is $9.35'$.

From the MPIfR survey Web-site mentioned above it is possible to
download both ``background'' and ``source'' total power maps,
which correspond to large and small-scale components of the total
intensity, as well as polarization maps. To derive the total Galactic
emission we have subtracted from the sum of ``background'' and ``source''
maps the isotropic component due to the CMB and to the contribution
of unresolved extragalactic sources. We have adopted, for this component,
$T_{A,{\rm isot}}=2.9\,$K.

\begin{figure}
\resizebox{\hsize}{!}{\includegraphics{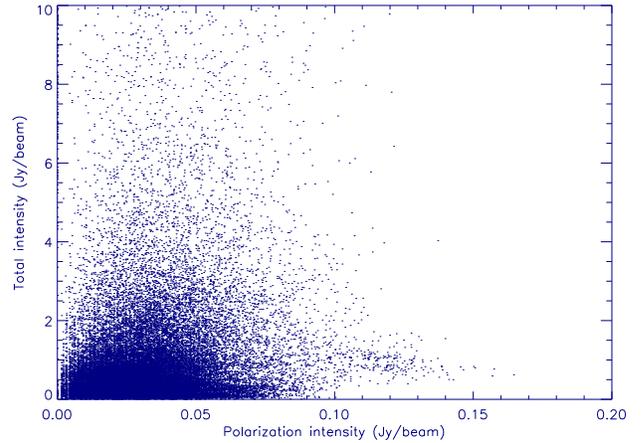}}
\caption{Total versus polarized intensity in the
$300^{\circ}\le {\rm l} \le 320^{\circ}$ region from the
D97 data at 2.4 GHz, showing no correlation at all.}
\label{fig1}
\end{figure}

\section{Analysis of survey data}

\subsection{Depolarization}

A comparison of total power and polarized  emission maps (Duncan \etal~1995,
1997; Uyaniker \etal~1998, 1999) shows little correlation. While the
total intensity clearly peaks on the Galactic Plane (apart from a number
of spurs and loops extending to high Galactic latitude), the polarized
intensity is much more uniformly distributed. Also, many sources which
are very intense in total power are not seen in polarized emission and,
conversely, bright regions of extended polarization do not appear to be
connected with sources of total-power emission (D97, U99).
This is seen in Fig.~1, showing, as an example, the
polarization intensity $\sqrt{Q^{2}+U^{2}}$ versus the total intensity for
the Parkes survey data in the region $300^{\circ}\le {\rm l}\le 320^{\circ}$
and $|b|\le 5^{\circ}$.

Two main factors may contribute to this situation. On one side, many bright
structures on the Galactic plane are (unpolarized) H{\sc ii} regions and
significant thermal radio emission is also expected between and above bright
H{\sc ii} regions in our Galaxy: Duncan \etal~(1995) estimate a thermal flux
level of the order of 100$\,$mJy per beam area at 2.4 GHz, comparable to
the level of the radio continuum often observed near these regions. The
large thermal contributions to the Galactic emission, which are
concentrated close to the Galactic plane, obviously make it
very unlikely that the total intensity power spectra for these regions can
be representative of the synchrotron power spectra at high
Galactic latitudes.

On the other side, differential Faraday rotation or variations of the magnetic
field orientation may strongly depolarize the emission from distant regions
of the Galaxy (Burn 1966, Gardner \& Whiteoak 1966; for a recent, detailed
discussion of depolarization mechanisms, see Sokoloff \etal~1998)
so that only the polarized emission of relatively local origin can be 
observed. The two factors may act together: variations in the density 
of thermally emitting electrons may lead to a large enough differential 
Faraday rotation to produce substantial depolarization; in addition, 
the magnetic field may be tangled by turbulent motions of the ionized gas, 
leading to further depolarization.

The Faraday rotation of the polarization position angle of a linearly 
polarized wave at a wavelength $\lambda$ traversing an ionized medium with
electron density $n_e$ and a regular magnetic field ${\bf B}$ 
is given by:

\begin{equation}
\phi(\lambda)=\hbox{RM}\lambda^2 \ \hbox{rad},
\label{RM1}
\end{equation}

where the rotation measure RM is the line-of-sight integral

\begin{eqnarray}
\hbox{RM} & = & {e^3\over 2\pi m_e^2c^4}\int n_e{\bf B}\cdot d{\bf l}= \nonumber \\
& = & 810 \int n_e(\hbox{cm}^{-3}) B_\parallel(\mu\hbox{G}) \cdot dl(\hbox{kpc})\ \hbox{rad}\,\hbox{m}^{-2} .
\label{RM2}
\end{eqnarray}

Since $\phi$ scales as $\nu^{-2}$, the Faraday rotation
is likely irrelevant at microwaves, where MAP and {\sc Planck} instruments will
operate, while it may strongly distort the power spectrum of polarized
emission at the decimetric wavelengths considered here, on one side
by wiping out polarized emission from far regions (due to differential 
rotation within the beamwidth or the bandwidth of observations) and, on 
the other side, introducing structure on a variety of scales 
(larger than the beamwidth) due to fluctuations in the thermal electron 
density distribution and/or magnetic field variations (in strength and/or 
direction) along the line of sight.

If the synchrotron emission arises throughout the depth of the Faraday
rotating medium, the polarization degree is reduced from the intrinsic
value $P_0$ to (Burn 1966):

\begin{equation}
P(\lambda)= P_0 {\sin\phi \over \phi} \ .
\label{depol}
\end{equation}

For $\phi = 1\,$rad, corresponding to $\hbox{RM}(\hbox{rad}\,\hbox{m}^{-2})
= 21.8$, 64.1, and 80.8 for $\nu = 1.4$, 2.4, and $2.7\,$GHz, respectively,
the depolarization amounts to about 16\%. Spoelstra (1984), from his analysis
of multifrequency data on linear polarization, found that the distribution
of RM over the sky shows a complex pattern with typical values (in the
region covered by Leiden surveys) of $8\,\hbox{rad}\,\hbox{m}^{-2}$.
As pointed out by the referee, however, Spoelstra's estimates of RMs
are actually lower limits because there may be field reversals along the line
of sight, his surveys have different resolutions and differential Faraday
depolarization can lead to small observed RMs (see Sokoloff \etal~1998).
Accurate estimates of the amount of depolarization were obtained, for the
Leiden survey at 820 MHz, by Berkhuijsen (1971) who determined the percentage
polarization using the absolutely calibrated total power survey at that
frequency. The polarization percentages given in that work should be
scaled by 0.8 (Berkhuijsen 1975) and are not corrected for the thermal
contribution from ionized gas, which decreases the polarization degree.

The implied RMs are relatively small over the surveyed area, particularly at
high Galactic latitudes.

The Galactic RM distribution is also probed by Faraday rotation
measurements towards pulsars and extragalactic radio sources. Since
pulsar distances can often be independently derived, pulsar data also
provide information on the variation of RM along the line of sight;
also, they appear to have no intrinsic Faraday rotation and hence
their observed RM arises entirely along the path to the observer.
Extragalactic sources can provide information on the Galactic medium out to
large distances, beyond those where pulsars are found; on the other hand,
extragalactic sources may have their own Faraday rotation which adds to
the Galactic contribution.

A catalogue of known pulsars, including values of RM, has been published
by Taylor \etal~(1993); an updated version is available
on the WEB (see Appendix of Taylor \etal~1993).
Additional RMs have been published by Manchester
\& Johnston (1995), Navarro \etal~(1997) and Han \etal~(1999); on the
whole, we have collected RMs for 318 pulsars.

Rotation measures for 674 extragalactic sources have been catalogued by
Broten \etal~(1988). Additional RMs have been published by Clegg \etal~
(1992).

An analysis of the RM of pulsars and extragalactic sources located in the sky
areas covered by polarization surveys considered here reveals a number of
regions where RMs can produce only a small Faraday depolarization ($\phi
\leq 1\,$rad).

One of these is the area $140^\circ \leq {\rm l} \leq 153^\circ$,
$3^\circ.7 \leq b \leq 10^\circ$, surveyed by U99. This
partly covers the highly polarized region referred to as the ``fan
region'' where the rotation measures have long been known to be small
(Bingham \& Shakeshaft 1967) and the magnetic field direction has to
be basically perpendicular to the line of sight.

















%





%






%

\begin{figure}
\resizebox{\hsize}{!}{\includegraphics{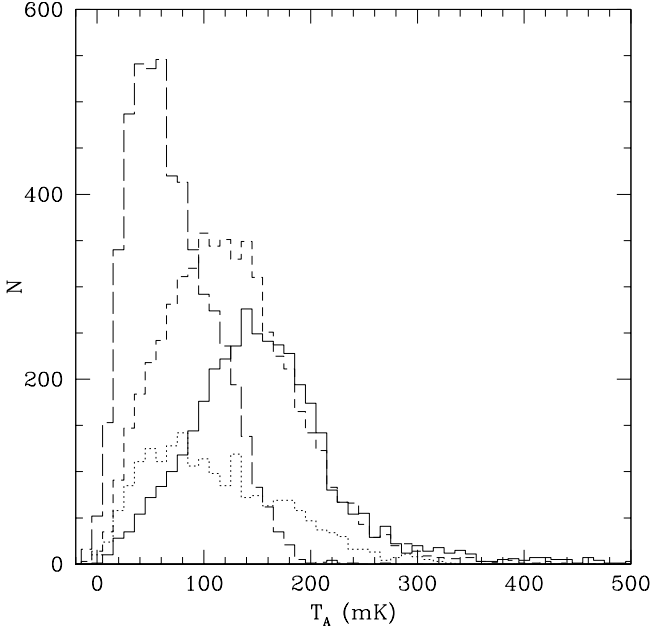}}
\caption{Distributions of polarized intensities for the U99 regions
centered at ${\rm l}= 50^\circ$ (solid line), $80^\circ$ ($b\geq 8^\circ$;
short dashes), $146^\circ.4$ (dotted line), and $200^\circ$ (long dashes). }
\label{fig2}
\end{figure}

\begin{figure}
\resizebox{\hsize}{!}{\includegraphics{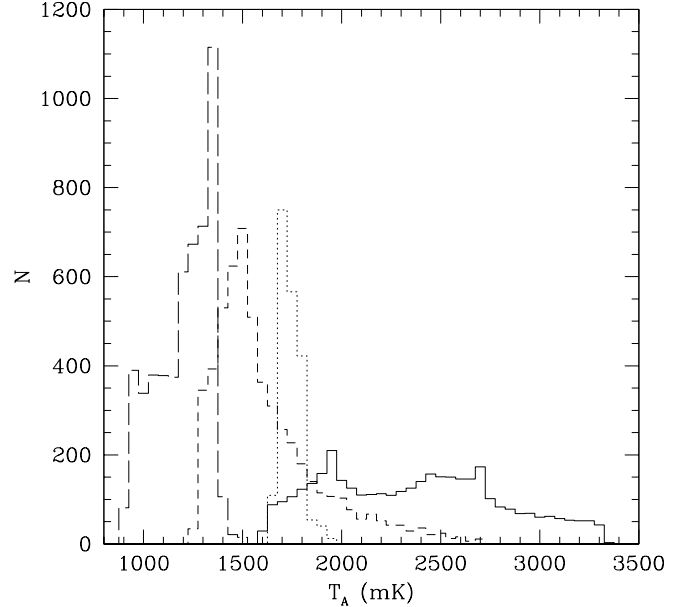}}
\caption{Distributions of total intensities (isotropic component
subtracted, see text) for the U99 regions
centered at ${\rm l}= 50^\circ$ (solid line), $80^\circ$ ($b\geq 8^\circ$;
short dashes), $146^\circ.4$ (dotted line), and $200^\circ$ (long dashes).
The maps used for this analysis do not include $\geq 5\sigma$ 
unresolved discrete sources; 
addition of these enhances the high intensity tails. }
\label{fig3}
\end{figure}

%










%

%










%
















\begin{table}

\caption{Polarization degrees of non-thermal emission at 1.4 GHz
for several U99 fields (Berkhuijsen, private communication).}

\begin{tabular}{rrr} \hline
l (deg) & b (deg) & $P (\%)$\\  
\hline
151.5            &  $+6$ &  20 \\
146\phantom{.0}  &  $+7$ &   4 \\
 50\phantom{.0}  & $+10$ &   3 \\
 78\phantom{.0}  & $+10$ &  19 \\
 86\phantom{.0}  & $+10$ &  12 \\
 80\phantom{.0}  & $-10$ &   9 \\
 92\phantom{.0}  & $-10$ &  10 \\ 
\hline
\label{Berk}
\end{tabular}

\end{table}

Berkhuijsen (private communication) has estimated the observed
polarization degree of non-thermal emission for several U99 fields discussed
in this paper (see Table~\ref{Berk}).
The total synchrotron emission was derived subtracting from the observed
intensities the contributions from extragalactic radio sources and the CMB
(3 K at 1.4 GHz) and from thermal emission. The fraction of thermal emission
at 1.4 GHz, $f_{\rm therm}$, was obtained from

\begin{equation}
f_{\rm therm}={1-(\nu_1/\nu_2)^{s-\gamma} \over 1-(\nu_1/\nu_2)^{t-\gamma}}
\end{equation}

{where $s$ is the observed spectral index (in terms of antenna temperature:
$T_A \propto \nu^{-s}$)
between $\nu_1=1420\,$MHz and $\nu_2=408\,$MHz taken, for each field, from
Reich \& Reich (1988), and $\gamma=2.9$, $t=2.1$ are the spectral
indices of non-thermal and thermal emission, respectively. The highest
polarization degree of non-thermal emission found in these fields is about
20\%, well below the theoretical maximum
corresponding to a uniform magnetic field (Ginzburg \& Syrovatskii 1965):}

\begin{equation}
\Pi = {3-3\gamma \over 1-3\gamma} \ ,
\label{depol}
\end{equation}

which, for $\gamma = 2.9$, is $\Pi = 74\%$. To depolarize to the observed
level with
differential Faraday rotation, $\hbox{RM} > 53~\hbox{rad}\,\hbox{m}^{-2}$
would be required, much higher than the RMs directly measured. Thus, other
depolarization mechanisms must be operating.

\begin{table*}
\caption{Moments of the distributions of total and polarized intensities.}
\begin{tabular}{lrrrrcccc} 
\hline
       & \multicolumn{2}{c}{mean (mK)} & \multicolumn{2}{c}{$\sigma$ (mK)} &
                         \multicolumn{2}{c}{skewness} &
\multicolumn{2}{c}{kurtosis} \\

           &  int & pol &  int &  pol &     int     &
pol        &  int & pol \\  \hline

Uyaniker ${\rm l}=50^\circ$  & 2480 & 155 &  550 &   70 &
$0.9 \pm 0.5$ & $1.3 \pm 0.1$ &$5.5\pm 3.5$ &$3.2\pm 0.3$ \\
\qquad no sources  & 2360 &  &  440 &    &
$0.051 \pm 0.007$ &  &$-0.89\pm 0.02$ &  \\
Uyaniker ${\rm l}=80^\circ${}$^\ast$  &  1640 &  125 &  310 &
60 &$19 \pm 12$&$0.54 \pm 0.05$  & $74 \pm 36$  & $1.0 \pm 0.1$ \\
\qquad no sources  &  1610 &  &  270 &   &$2.0 \pm 0.06$&  & $2.0 \pm 0.1$ &  \\
Uyaniker ${\rm l}=146^\circ.4$  & 1840 & 110 &  190 &
65 &$28\pm 7$ & $0.43\pm 0.04$& $58 \pm 10$ & $-0.07 \pm 0.08$ \\
\qquad no sources  & 1740 &  &  50 &  &$0.67\pm 0.07$ & & $1.0 \pm 0.1$ &  \\
Uyaniker ${\rm l}=200^\circ$  &  1270 &  70 &  230 &   40 &
$23 \pm 11$   & $0.36 \pm 0.03$ &$82 \pm 22$ & $-0.05 \pm 0.1$ \\
\qquad no sources  &  1200 &  &  140 &  & $0.23 \pm 0.01$   &  &$-0.94 \pm 0.03$ &
 \\
D97 $240^\circ \leq {\rm l} \leq 260^\circ$  &  161 &  13 &  110 &    9 &$60 \pm 40$ &
$0.26 \pm 0.02$ &$210 \pm 80$ & $0.08 \pm 0.07$ \\
\qquad no sources                             &  129 &     &   70 &      &
$0.004\pm 0.003$ &          &$-0.15 \pm 0.03$ &  \\
D97 $280^\circ \leq {\rm l} \leq 300^\circ$  &  425 & 106 & 1741 &   33  &$340\pm 40$   &
$0.98 \pm 0.05$ &$470 \pm 60$ & $1.0 \pm 0.1$ \\
\qquad no sources                             &  192 &     &  227 &      &
$1.41 \pm 0.04$ &        & $0.53\pm 0.07$ &  \\
D97 $300^\circ \leq {\rm l} \leq 320^\circ$  &  577 &  15 &  930 &   15 &
$90 \pm 20$   & $0.88\pm  0.05$ & $200\pm 30$  & $0.7\pm 0.2 $ \\
\qquad no sources         &  423 &      & 404 &   & $0.59\pm 0.02$&  &
$-0.22\pm 0.04$ &  \\
D99 $20^\circ \leq {\rm l} \leq 30^\circ$    &  317 &   32 & 907 & 20 &$130\pm 40$ &
$1.3\pm 0.3$ &$270\pm 70$ & $4\pm 2$  \\
D99 $55^\circ \leq {\rm l} \leq 65^\circ$  & 72 & 32 & 18 & 11 &$1500\pm 300$ &
$0.49\pm 0.03$ &$2400\pm 300$ & $0.6\pm 0.1$  \\ \hline
\label{moments}
\end{tabular}

  \  {}$^\ast$ Analysis limited to the region $8^\circ \leq b \leq 15^\circ$
\end{table*}

\begin{figure}
\resizebox{\hsize}{!}{\includegraphics{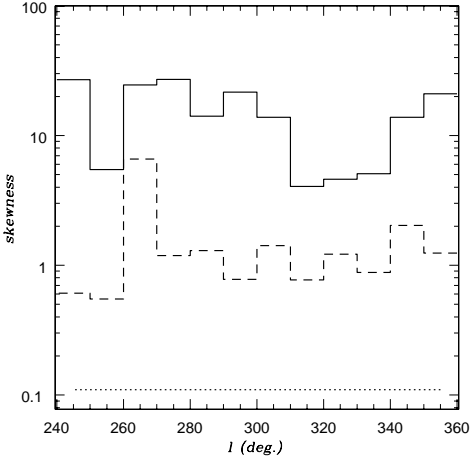}}
\vskip-45pt
\caption{Global skewness of the total (upper, solid line) and polarized
emissions (dashed) for the D97 survey compared with the rms
error on the skewness derived from Monte Carlo simulations for a
$10^\circ\times 10^\circ$ patch (grid $4'\times 4'$ as default in the 
D97 data) assuming the same power spectrum for the
signals but a Gaussian distribution (straight dotted line at the bottom).}
\label{fig3'}
\end{figure}

\subsection{Distribution of total intensity and polarization signals}

All maps were put onto the same $12'\times 12'$ grid. Although these
cells are not totally independent (the telescope beamwidth is $9.35'$),
we do not need to worry much about that since the derived
power spectrum is affected only on small scales, where fluctuations
due to point sources dominate anyway (De Zotti \etal~1999). 

It is clear from Figs.~\ref{fig2} and \ref{fig3} that the distributions of
both total and polarized emissions are distinctly non-Gaussian.
The statistical significance of this visual impression can be 
quantified by computing the moments of order $>2$:

\begin{equation}
\mu_{\rm i} = {\sum_{j=1}^{n}(S_{\rm j} - \bar{S})^{\rm i}\over n - 1}\ ,
\label{momenti}
\end{equation}

where $S_{\rm j}$ is the measured signal (polarized or unpolarized 
intensity in the j-th cell) and $\bar{S}$ is its mean over the 
considered sky region.

It is convenient to define the dimensionless
quantities $\beta_1=\mu_3^2/\mu_2^{3}$, $\beta_2=\mu_4/\mu_2^{2}$,
$\beta_3=\mu_3\mu_5/\mu_2^{4}$, $\beta_4=\mu_6/\mu_2^{3}$,
$\beta_5=\mu_7 \mu_3/\mu_2^{5}$, $\beta_6=\mu_8/\mu_2^{4}$.

Usual definitions of skewness and of kurtosis are
${\rm skewn} = \sqrt{\beta_1}$ and ${\rm kurt} = \beta_2-3$,
which vanish for a Gaussian distribution. The probable errors of
$\beta_1$ and $\beta_2$ are given by (Pearson 1924):

\begin{eqnarray}
\Delta\beta_{1} &=& 0.6745 (\beta_1 /n)^{1/2}\cdot\nonumber\\
& \cdot &{\left(4\beta_4-24\beta_2+36+9\beta_1\beta_2 
-12\beta_3+35\beta_1\right)^{1/2}}
\end{eqnarray}

\begin{eqnarray}
\Delta\beta_{2} &=& 0.6745 n^{-1/2}\cdot\nonumber\\
\cdot (\beta_6 &-& 4\beta_2\beta_4+4\beta_2^{3}-\beta_2^{2}+
16\beta_1\beta_2-8\beta_3+16\beta_1)^{1/2}.
\label{errbeta}
\end{eqnarray}

The values of skewness and kurtosis for the regions we have analyzed, and
their errors, are given in Table~\ref{moments}, together with the mean total
and polarization intensities and their standard deviations $\sigma$.
Note that, for the U99 area centered at ${\rm l}=80^\circ$, we have
limited our analysis to the region above $b =8^\circ$ to avoid the relatively
large hole in the map at lower Galactic latitudes.

The deviations from the Gaussian value (zero) of skewness and/or kurtosis
are in general highly significant, both for total and for polarized emissions.
This conclusion is strengthened by the Monte Carlo simulations reported in
Fig.~\ref{fig3'} for $10^{\circ}\times 10^{\circ}$ patches (grid 
$4'\times 4'$ as default in the D97 data), 
showing that the skewness differs from $0$ at a
$\gsim 10\sigma$ level in the case of polarization and even more for total
intensity. This may be a difficulty for methods
involving Wiener filtering of the data to remove foreground contributions
to CMB maps (Bouchet \etal~1999) since a Gaussian approximation for the
distribution of foreground signals is assumed. On the other hand, Independent
Component Analysis algorithms (Baccigalupi \etal~2000) require that
all independent components contributing to the observed maps, except, at most,
one have non-Gaussian distributions.

\begin{figure}
\resizebox{\hsize}{!}{\includegraphics{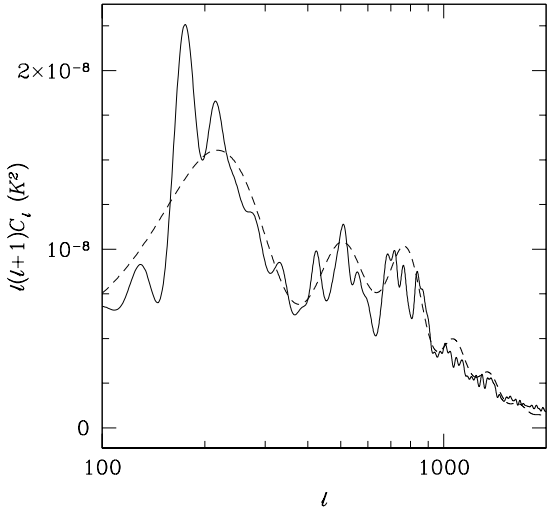}}
\vskip-15pt
\caption{Reconstructed power spectrum of CMB temperature fluctuations
from a simulated $10^\circ\times 10^\circ$ patch (solid line) compared
with the theoretical spectrum predicted by a standard CDM model,
averaged over the whole sky (dashed line).}
\label{f1spectra}
\end{figure}

\section{Intensity and polarization power spectra}

\label{intensity}

It is currently standard practice to adopt as a
statistical measure of the temperature pattern on the celestial
sphere, the power spectrum of temperature fluctuations, $C_{\ell}$,
which turned out to be of fundamental importance in studies of
CMB anisotropies.

The $C_{\ell}$'s are defined as follows.
The spherical harmonic expansion of the sky signal $s$ writes:

\begin{equation}
s(\hat{n})=\sum_{\ell m}a_{\ell m}Y_{\ell}^{m}(\hat{n})\ ,
\label{ylm}
\end{equation}

where $a_{\ell m}$ are the expansion coefficients,
$\hat{n}$ is the unit vector in the direction $(\vartheta ,\phi )$, 
and the spherical harmonic $Y_{\ell}^{m}$ is related to the Legendre
function, $P_{\ell}^{|m|}$, of degree $\ell$ and order $m$ ($|m|\leq \ell$) by
$Y_{\ell}^{m}(\vartheta ,\phi ) = c_{\ell}^{m} P_{\ell}^{|m|}(\cos\vartheta)
\exp(i m\phi)$; the coefficients $c_{\ell}^{m}$ normalize the spherical
harmonics to $\int d\Omega Y_{\ell}^{m}  Y_{\ell'}^{m'} = \delta_{\ell\ell'}
\delta{mm'}$. The correlation function $C(\theta )=
<s(\hat{n})s(\hat{n}')>_{\hat{n}\cdot{\hat{n}'}=\cos\theta}$
can be expanded into Legendre polynomials,
with coefficients given by

\begin{equation}
C_{\ell}={1\over 2\ell +1}\sum_{m=-\ell}^{\ell}|a_{\ell m}|^{2}\ .
\label{cl}
\end{equation}

The multipole $\ell$ corresponds to the angular scale

\begin{equation}
\theta \simeq 180/\ell\ {\rm degrees}\ .
\label{ltheta}
\end{equation}

We have projected each sky patch onto a null signal sphere
and computed the power spectrum using of the HEALPix
tools (G\'orski \etal~1998).
The coefficients were renormalized by simply dividing
by the fractional coverage of the sky.

Note that the use of spherical harmonics is not strictly necessary, given
that we are dealing with limited areas of the sky. In fact, for the common
data sets, our results agree with those by Tucci \etal~(2000) who
resorted to a standard Fourier analysis technique.
We preferred, however, to stick to the spherical harmonic analysis to
be used for the all sky MAP and {\sc Planck} data.

We have tested our method using a simulation of a purely 
cosmological signal.
We have generated an all sky map of the CMB temperature distribution
as predicted by a standard model for a Cold Dark Matter (CDM)
dominated universe, at a resolution of about $3.5'$. We have then applied
our algorithm to a randomly chosen $10^{\circ}\times 10^{\circ}$ patch.
In Fig.~\ref{f1spectra} we show the recovered power spectrum for the patch
compared with the theoretical one (which is, of course, an all-sky average).
The reconstructed spectrum oscillates around the theoretical one, due
to the sample variance, \ie to the very limited sampling
of a random, all sky process. However, the main features of the spectrum
are recovered.

Linear polarization is described by the Stokes parameters $Q$ and $U$,
from which the total polarization intensity $PI=\sqrt{Q^{2}+U^{2}}$ and
the polarization angle $\phi_{P}=\arctan{(U/Q)}$,
can be derived. $PI$ is a scalar quantity
that can be expanded into spherical harmonics [Eq.~(\ref{ylm})]; its
power spectrum coefficients $C_{\ell}^{P}$ can be computed as in
Eq.~(\ref{cl}).

In order to derive the true power spectrum coefficients from the
measured ones, $C_{\ell}^{\rm map}$, we must allow for the contribution
of instrumental noise and for the effect of the detector response
function $b(\ell)$:

\begin{equation}
\label{bestcl}
C_{\ell}={C_{\ell}^{\rm map}-C_{\ell}^{\rm noise}\over W(\ell )}\ ,
\ \ W(\ell )=b(\ell )^{2}\ .
\end{equation}

The beam response for each multipole is modelled by
\begin{equation}
b(\ell )=\exp [-\ell (\ell +1)\sigma_{\rm beam}^{2}/2]\ .
\label{wl}
\end{equation}

where $\sigma_{\rm beam} = {\rm FWHM}/(2\sqrt{2\ln2})$ and FWHM is the
full width at half maximum, in radians. In the case of D97, the beam
is slightly elliptical and Eq.~(\ref{wl}) has been modified accordingly.

The instrumental noise has an approximately flat power spectrum:
\begin{equation}
C_{\ell}^{\rm noise}=4\pi{\sigma^{2}_{\rm noise}\over N_{\rm pixels}}\ .
\label{clnoise}
\end{equation}

For the surveys of D97, D99 and U99 we have adopted the values of the
instrumental noise given by the authors (taking into account the
different sensitivities reached by D97 in different regions).
In the case of the BS76 data, we found it convenient
to treat the noise as a parameter to be determined (see subsection 4.2).

We have analyzed $10^\circ\times 10^\circ$ patches of the surveys by D97,
D99 and U99. Correspondingly, the minimum value of $\ell$ for which the power
spectrum can be estimated is $\simeq 100$, corresponding to an angular scale
of about $2^\circ$: as the angular scale approaches the size of the patch,
the effects of poor sampling become unacceptably large. Information on
larger angular scales is provided by the BS76 data. The maximum value
of $\ell$ is determined by the angular resolution of the survey; we have
$\ell_{\rm max} \simeq 800$.

\begin{figure}
\resizebox{\hsize}{!}{\includegraphics{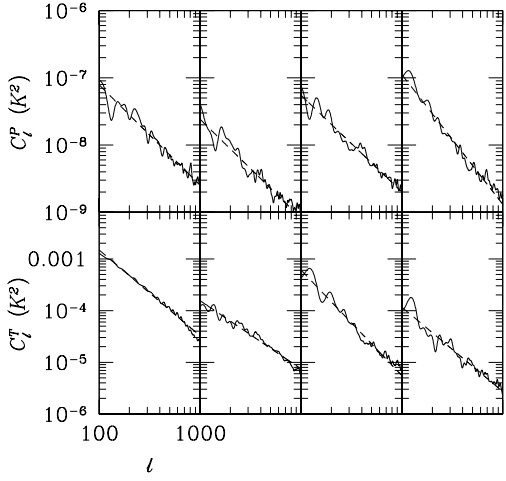}}
\vskip-45pt
\caption{Polarization (upper panels) and total intensity (lower panels) angular
power spectra for the D97 region $360^{\circ}\ge {\rm l}\ge 320^{\circ}$.
The panels correspond to $10^\circ$ intervals in longitude, with l decreasing
from left to right.
The dashed lines represent power law fits.}
\label{f4spectra}
\end{figure}

\begin{figure}
\resizebox{\hsize}{!}{\includegraphics{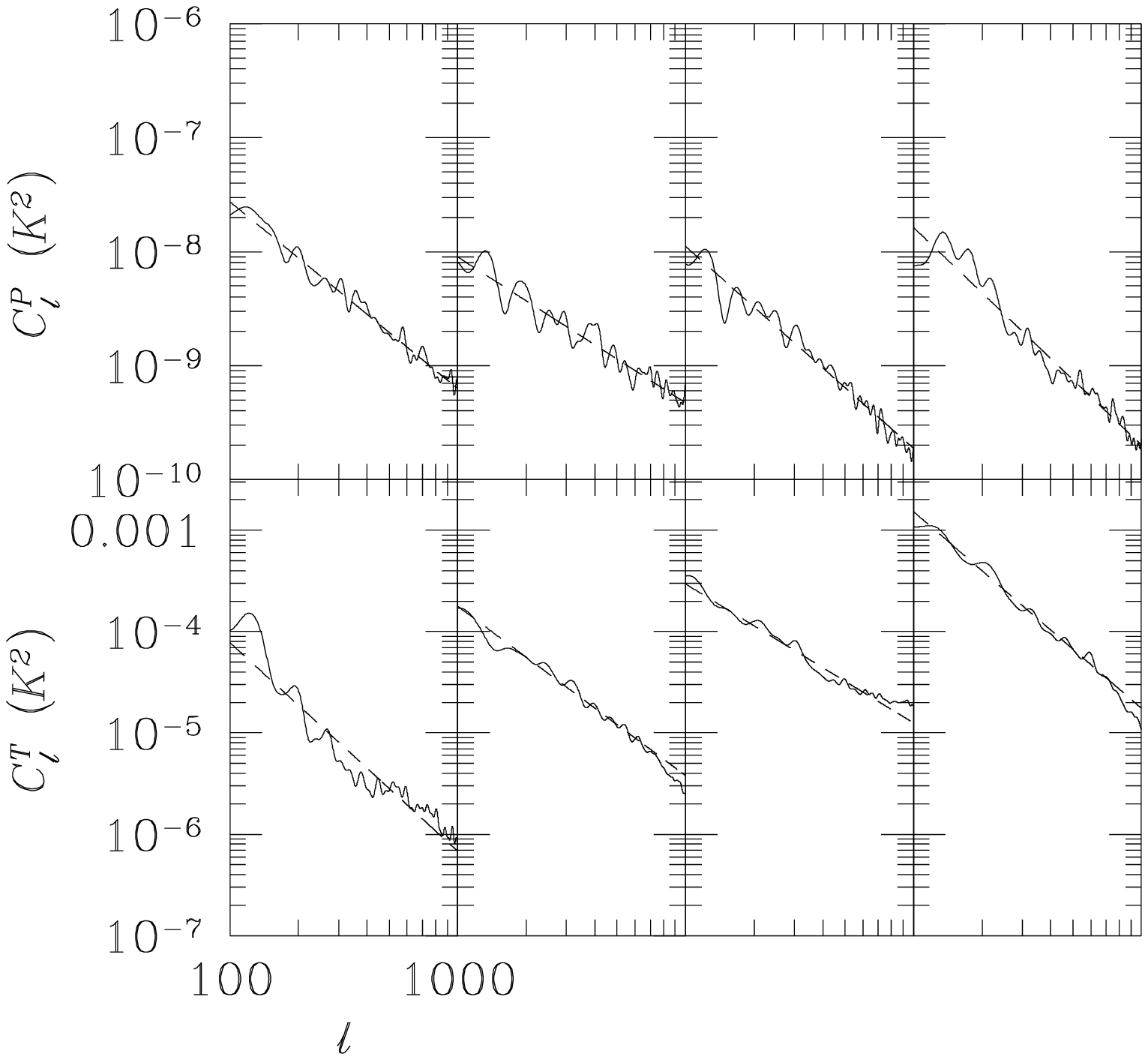}}
\vskip-45pt
\caption{Polarization (upper panels) and total intensity (lower panels) angular
power spectra for the D97 region $320^{\circ}\ge {\rm l}\ge 280^{\circ}$.
The panels correspond to $10^\circ$ intervals in longitude, with l decreasing
from left to right. The dashed lines represent power law fits.}
\label{f5spectra}
\end{figure}

\begin{figure}
\resizebox{\hsize}{!}{\includegraphics{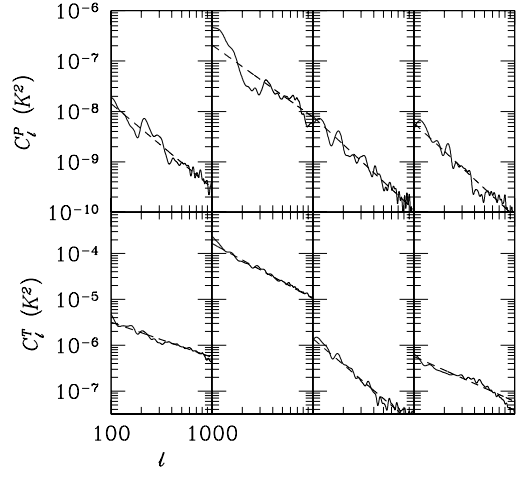}}
\vskip-45pt
\caption{Polarization (upper panels) and total intensity (lower panels) angular
power spectra for the D97 region $280^{\circ}\ge {\rm l}\ge 240^{\circ}$.
The panels correspond to $10^\circ$ intervals in longitude, with l decreasing
from left to right. The dashed lines represent power law fits.}
\label{f6spectra}
\end{figure}

\subsection{Low and medium Galactic latitudes}

We have focused our analysis on the low RM regions of the D97, D99, and U99
surveys. As detailed in the following, we find that these regions do possess
remarkably similar polarization power spectra.

Following Maino \etal~(1999) we divided the D97 data in twelve
$10^{\circ}\times 10^{\circ}$ subpatches, and we evaluated the
power spectrum of both total intensity and polarization for each patch,
over the range $100\le\ell\le 800$.
The results are plotted in Figs. \ref{f4spectra}--\ref{f6spectra}.
In each panel, the dashed line shows a power law fit
($C_{\ell}=\alpha\cdot \ell^{\beta}$).

\begin{figure}
\resizebox{\hsize}{!}{\includegraphics{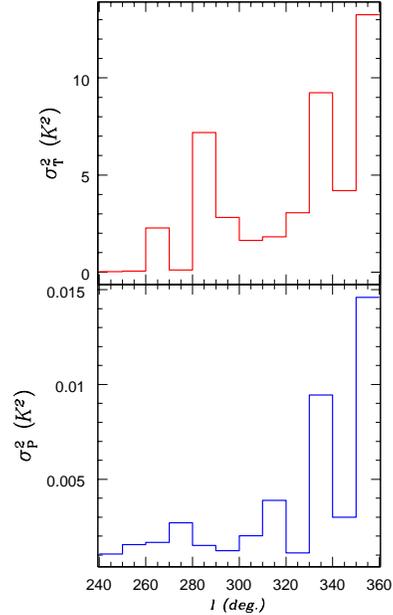}}
\caption{Total (upper panel) and polarized intensity fluctuations
in the area surveyed by D97.}
\label{f2spectra}
\end{figure}

\begin{figure}
\resizebox{\hsize}{!}{\includegraphics{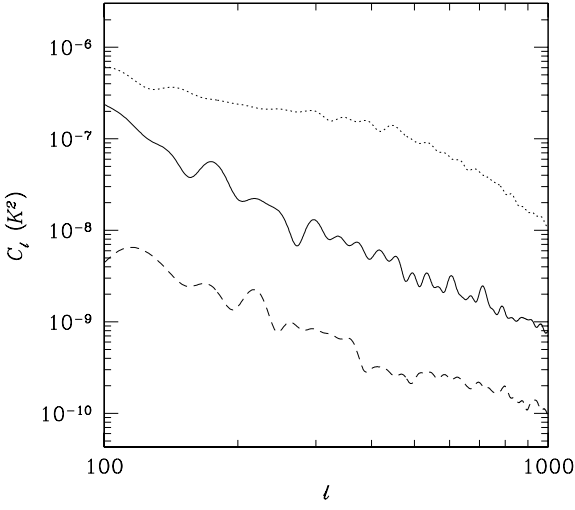}}
\caption{Power spectrum for total intensity (dotted line)
and polarization (dashed line) in the
$240^{\circ}\le {\rm l}\le 250^{\circ}$ D97 region.
The solid line shows the power spectrum coefficients $C_{\ell}$ of total
intensity before (dotted line) and after (solid line) removing the
most prominent sources as described in the text, in comparison with
polarization  (dashed line).}
\label{f3spectra}
\end{figure}

\begin{table}

\caption{Bright H{\sc ii} regions between ${\rm l} =300^{\circ}$ and
${\rm l} =320^{\circ}$.}
\begin{tabular}{lrrrccc} 
\hline
l (deg.) & $b$ (deg.) & $S_{2.7{\rm GHz}}\,$(Jy) & Size ($'$) \\ \hline
305.1 & 0.1 & 16.3  &  5 \\
305.2 & 0.0 & 62.2  &  7  \\
305.2 & 0.2 & 50.1  &  4.8 \\
305.4 & 0.2 & 62.2  &  3.5 \\
305.6 & 0.0 & 37.6  &  8   \\
307.6 & -0.3 & 12.2 &  4.2  \\
308.6 & 0.6 & 17.8  &  7.8  \\
308.7 & 0.1 & 12.0  &  --  \\
308.7 & 0.6 & 21.0  &  --  \\
309.6 & 1.7 & 51.0  &  7.5 \\
310.8 & -0.4 & 16.6 & 10.9\\
311.9 & 0.1 & 12.9  &  4  \\
311.9 & 0.2 & 11.5  &  4  \\
312.3 & -0.3 & 10.0 &  --  \\
316.3 & 0.0 & 12.0  &  --  \\
316.8 & -0.1 & 43.2 &  2.7 \\
317.0 & 0.3 & 15.0  &  7   \\
318.0 & -0.8 & 11.0 & 10.8 \\
319.2 & -0.4 & 12.4 &  5.9 \\
320.2 & 0.8 & 11.0  &  1.8 \\
320.3 & -1.4 & 12.0 &  --  \\
320.3 & -1.0 & 23.0 &  --  \\
320.4 & -1.1 & 17.5 &  6   \\
320.4 & -1.0 & 13.0 &  5.6  \\ 
\hline
\end{tabular}
\label{Hii}
\end{table}

Fig.~\ref{f2spectra} shows the fluctuations around the mean of both
total and polarized intensities for the D97 survey, as a function
of the angular distance from the Galactic center, for regions 
$10^\circ$ wide in longitude.
It may be noted that polarization and total intensity fluctuations are
correlated only near the Galactic center. The intensity peaks are associated
to the most prominent Galactic sources, i.e. the Galactic center and
the Vela Supernova remnant ($260^{\circ}\le {\rm l} \le 270^{\circ}$).
For polarization, the signals at $320^{\circ}\le {\rm l} \le 360^{\circ}$
are due to features appearing near the Galactic center
such as the polarization ``plume" (Duncan \etal~1998), while
away from the Galactic center, at ${\rm l} \le 320^{\circ}$,
only the supernova remnant produces peaks both in total and in polarized
intensity.

Also, and most important, this figure shows that the total intensity
fluctuations are higher by 2 or 3 orders of magnitude than the polarization 
ones, implying a polarization degree
much lower than the maximum expected for undepolarized
synchrotron emission.

In fact, most of the observed intensity close to the Galactic plane appears
to come from bright H{\sc ii} regions. For example, the region between
${\rm l} =300^{\circ}$ and ${\rm l} =320^{\circ}$
contains 95 catalogued bright H{\sc ii} regions
(Paladini \etal~2000); the brightest ones are listed in Table \ref{Hii} with
their flux density at 2.7 GHz and their angular sizes which are
typically in the range $2'$--$10'$. It is easily checked that these sources
account for most, if not all, of the total intensity reported by D97.
The intensity fluctuations due to them can be roughly estimated to be:

\begin{equation}
\sigma_{HII}^{2}\simeq {\sum_{i}S_{i}^{2}\over N_{\rm pixels}}\ ,
\label{rmsHii}
\end{equation}

where $S_{i}$ is the flux of the $i-th$ H{\sc ii} region in the
catalogue, scaled to the D97 resolution and to 2.4 GHz by assuming a
spectral index of 0.1 ($S_{\nu} \propto \nu^{-0.1}$), appropriate for the
free-free emission. This yields $\sigma_{H{\sc ii}}\simeq 1\ {\rm Jy/beam}$,
close to the value of 1.4 Jy/beam that we obtain from the D97 data.

The free-free emission from H{\sc ii} regions is not polarized, but Thomson
scattering by the electrons in the H{\sc ii} region itself may polarize the
radiation tangentially to the edges of the cloud structure at a maximum 
level of about 10\% for an optically thick cloud (Keating \etal~1998).














Another indication that sources contributing most of the intensity in the
area surveyed by D97 are unpolarized is obtained by removing the highest
intensity peaks and replacing their signal with the median value in an
annulus around them. As shown by Fig.~\ref{f3spectra}, this has the effect
of strongly decreasing the amplitude of the total intensity power spectrum,
particularly at intermediate and small angular scales, making its shape
quite similar to that of the polarization power spectrum. Also, the
polarization degree becomes $\simeq 0.3$, consistent with moderately
depolarized synchrotron emission.

\begin{table}
\caption{Power spectrum parameters of total intensity
for the D97 survey. The first column gives the Galactic longitude of the
patch center.}

\begin{tabular}{l c c c c} 
\hline
l (deg.) & $\alpha\ ({\rm K}^{2})$ & $\beta$ &
$\Delta\alpha\ ({\rm K}^{2})$ & $\Delta\beta$\\ 
\hline
355 & 3.1 & -1.66 & 0.1 & 6$\cdot 10^{-3}$\\
345 & 6.8$\cdot 10^{-2}$ & -1.32 & 4$\cdot 10^{-3}$ & 1$\cdot 10^{-2}$\\
335 & 7.3 & -2.04 & 0.7 & 2$\cdot 10^{-2}$\\
325 & 0.21 & -1.63 & 2$\cdot 10^{-2}$ & 2$\cdot 10^{-2}$\\
315 & 1.0 & -2.06 & 0.2 & 3$\cdot 10^{-2}$\\
305 & 0.39 & -1.67 & 2$\cdot 10^{-2}$ & 1$\cdot 10^{-2}$\\
295 & 0.16 & -1.37 & 1$\cdot 10^{-2}$ & 1$\cdot 10^{-2}$\\
285 & 11.6 & -1.94 & 0.7 & 1$\cdot 10^{-2}$\\
275 & 1.1$\cdot 10^{-4}$ & -0.76 & 5$\cdot 10^{-6}$ & 7$\cdot 10^{-3}$\\
265 & 4.2$\cdot 10^{-2}$ & -1.19 & 2$\cdot 10^{-3}$ & 7$\cdot 10^{-3}$\\
255 & 7.6$\cdot 10^{-3}$ & -1.87 & 8$\cdot 10^{-4}$ & 2$\cdot 10^{-2}$\\
245 & 5.3$\cdot 10^{-5}$ & -0.98 & 4$\cdot 10^{-6}$ & 1$\cdot 10^{-2}$\\
\hline
\label{fitint}
\end{tabular}
\end{table}

\begin{table}
\caption{Power spectrum parameters of polarized intensity
for the D97 survey. The first column gives the Galactic longitude of the
patch center.}

\begin{tabular}{l c c c c} 
\hline
l (deg.) & $\alpha\ ({\rm K}^{2})$ & $\beta$ &
$\Delta\alpha\ ({\rm K}^{2})$ & $\Delta\beta$\\ \hline
355 & 7.5$\cdot 10^{-5}$ & -1.49 & 7$\cdot 10^{-6}$ & 2$\cdot 10^{-2}$\\
345 & 2.1$\cdot 10^{-5}$ & -1.47 & 2$\cdot 10^{-6}$ & 2$\cdot 10^{-2}$\\
335 & 4.2$\cdot 10^{-5}$ & -1.44 & 4$\cdot 10^{-6}$ & 1$\cdot 10^{-2}$\\
325 & 8.1$\cdot 10^{-4}$ & -1.93 & 8$\cdot 10^{-5}$ & 2$\cdot 10^{-2}$\\
315 & 5.2$\cdot 10^{-5}$ & -1.64 & 4$\cdot 10^{-6}$ & 1$\cdot 10^{-2}$\\
305 & 3.3$\cdot 10^{-6}$ & -1.28 & 4$\cdot 10^{-7}$ & 2$\cdot 10^{-2}$\\
295 & 3.9$\cdot 10^{-5}$ & -1.77 & 4$\cdot 10^{-6}$ & 2$\cdot 10^{-2}$\\
285 & 1.1$\cdot 10^{-4}$ & -1.91 & 1$\cdot 10^{-5}$ & 2$\cdot 10^{-2}$\\
275 & 3.1$\cdot 10^{-5}$ & -1.67 & 4$\cdot 10^{-6}$ & 2$\cdot 10^{-2}$\\
265 & 1.5$\cdot 10^{-4}$ & -1.43 & 3$\cdot 10^{-5}$ & 3$\cdot 10^{-2}$\\
250 & 4.0$\cdot 10^{-5}$ & -1.86 & 5$\cdot 10^{-6}$ & 2$\cdot 10^{-2}$\\
245 & 3.1$\cdot 10^{-5}$ & -1.86 & 5$\cdot 10^{-6}$ & 2$\cdot 10^{-2}$\\
\hline
\end{tabular}
\label{fitpol}
\end{table}

As already mentioned, the power spectra obtained for each patch have been
fitted with simple power laws:

\begin{equation}
C_{\ell}=\alpha\ell^{\beta}\qquad\qquad 100\le\ell\le 800 \ .
\label{fit}
\end{equation}

The values of the parameters $\alpha$ and $\beta$ have been obtained by
minimizing the quantity:

\begin{equation}
\sigma^{2}_{\rm fit}=
\sum_{\ell =\ell_{\rm min}}^{\ell =\ell_{\rm max}}
[\log_{10}C_{\ell}-\log_{10}\alpha -\beta\log_{10}\ell ]^{2}\ .
\label{fitsigma}
\end{equation}

The best-fit values of the parameters are listed in Tables~\ref{fitint}
(for total intensity) and \ref{fitpol} for polarization. The Tables also
give the errors $\Delta\alpha ,\Delta\beta$ calculated following
the standard prescriptions [see, e.g. Press \etal~(1996), Chapter 15], with

\begin{equation}
\Delta C_{\ell}^{2}=
\sum_{\ell=\ell_{\rm min}}^{\ell=\ell_{\rm max}}
{(\log_{10}C_{\ell}-\log_{10}\alpha-\beta\log_{10}\ell )^{2}
\over
{\ell_{\rm max}-\ell_{\rm min}-2}}\ .
\label{fitsignal}
\end{equation}

The visual impression that the fit is generally very good is quantitatively
confirmed by the fact that in all cases $\Delta\alpha\ll\alpha$ 
and $\Delta\beta\ll\beta$.

In the case of total intensity, the slope $\beta$ shows 
large variations (from $\simeq -0.8$ to $\simeq -2$) along 
the Galactic plane. Steeper slopes correspond to regions where 
diffuse emission dominates; point sources
add power on small scales (large $\ell$). The effect of point sources 
accounts for our finding of total intensity power spectra flatter (and, 
in some regions, much flatter) than derived from previous analyses of 
lower resolution surveys. In fact, analyses of the 408 MHz Haslam 
\etal~(1982) map (Tegmark \& Efstathiou 1996) and of the 1420 MHz 
Reich \& Reich (1988) map (Bouchet \& Gispert 1999) yielded values 
of $\beta \simeq -3$. Similar values of $\beta$ are also found from 
U99 ``background'' maps (\ie those obtained removing $\geq 5\sigma$ 
unresolved sources); in this case, we find $\beta \simeq -2.6$, $-2.8$, 
and $-3.35$ for the patches centered at ${\rm l}=50^\circ$, 
${\rm l}=146^\circ.4$ ${\rm l}=196^\circ.8$, respectively.
The situation is quite different for polarization. In general, the variations
of $\beta$ are substantially smaller. Also, there are several regions
(at ${\rm l}\le 320^{\circ}$, with exception of the Vela supernova remnant)
where the power spectra are remarkably similar and described,
for $100 \leq \ell \leq 800$, by:

\begin{eqnarray}
C^{P}_{\ell}&=&(1.1\pm 0.7)\cdot 10^{-9}\cdot
\left({\ell\over 450}\right)^{-1.7\pm 0.2}\cdot\nonumber\\
&\cdot&\left({\nu\over 2.4{\rm GHz}}\right)^{-5.8}\ {\rm K}^{2}\ ,
\label{fitduncan}
\end{eqnarray}

where we have explicitly indicated the typical frequency dependence of
Galactic synchrotron emission. We find essentially the same
power spectra also for several D99 regions and for the U99 regions with
low rotation measures (see below).
Therefore, we consider it as typical of the diffuse Galactic
polarized synchrotron emission.

The D99 survey at $2.7\,$GHz covers a region close to
the Galactic plane at $-5^{\circ}\le b\le 5^{\circ}$,
$4.9^{\circ}\le {\rm l}\le 74^{\circ}$. We selected five squared
$10^{\circ}\times 10^{\circ}$ patches centered at ${\rm l} =25^{\circ}$,
$35^{\circ}$, $45^{\circ}$, $55^{\circ}$, and $60^{\circ}$.

\begin{figure}
\resizebox{\hsize}{!}{\includegraphics{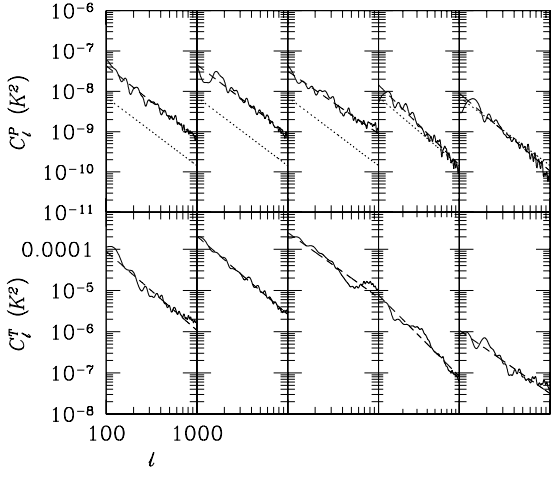}}
\vskip-40pt
\caption{Polarized (top) and total intensity (bottom) angular power
spectra for the D99 regions $20^{\circ}\ge {\rm l}\ge 30^{\circ}$,
$30^{\circ}\ge {\rm l}\ge 40^{\circ}$, $40^{\circ}\ge {\rm l}\ge 50^{\circ}$,
$50^{\circ}\ge {\rm l}\ge 55^{\circ}$, $55^{\circ}\ge {\rm l}\ge 65^{\circ}$
from left to right respectively. The dashed lines represent power law fits.
The dotted line represents Eq.~(18). For each panel, $\ell$ runs from 100 to
1000 on a logarithmic scale.}
\label{f7spectra}
\end{figure}

The angular power spectra are reported in Figure \ref{f7spectra},
and the best fit values for the coefficients of the power law are in Tables
\ref{fitintD99} and \ref{fitpolD99}.
The dotted lines plotted in the panels showing the polarization power spectra
represent Eq.~(\ref{fitduncan}). Clearly, the polarization power spectra
derived from the D99 survey are remarkably close to those found for
the D97 survey, scaled using a typical synchrotron spectral index of $2.9$.

%





%

\begin{table}
\caption{Power spectrum parameters for total intensity
in the $2.7\,$GHz D99 data. The first column gives the Galactic longitude
of the the patch center.}
\begin{tabular}{l c c c c} 
\hline
l (deg.) & $\alpha\ ({\rm K}^{2})$ & $\beta$ &
$\Delta\alpha\ ({\rm K}^{2})$ & $\Delta\beta$\\ \hline
\hline
25 & 0.62 & -1.92 & 7$\cdot 10^{-2}$ & 2$\cdot 10^{-2}$\\
35 & 1.32 & -1.90 & 7$\cdot 10^{-2}$ & 9$\cdot 10^{-3}$\\
45 & 0.37 & -1.58 & 5$\cdot 10^{-2}$ & 2$\cdot 10^{-2}$\\
55 & 6.7$\cdot 10^{-2}$ & -1.97 & 8$\cdot 10^{-3}$ & 2$\cdot 10^{-2}$\\
60 & 1.3$\cdot 10^{-3}$ & -1.54 & 2$\cdot 10^{-4}$ & 2$\cdot 10^{-2}$\\
\hline
\end{tabular}
\label{fitintD99}
\end{table}

\begin{table}
\caption{Power spectrum parameters of the polarized intensity
in the $2.7\,$GHz D99 data. The first column gives the Galactic longitude of
the patch center.}
\begin{tabular}{l c c c c} 
\hline
l (deg.) & $\alpha\ ({\rm K}^{2})$ & $\beta$ &
$\Delta\alpha\ ({\rm K}^{2})$ & $\Delta\beta$\\ \hline
25 & 1.6$\cdot 10^{-4}$ & -1.79 & 1$\cdot 10^{-5}$ & 1$\cdot 10^{-2}$\\
35 & 1.2$\cdot 10^{-4}$ & -1.72 & 1$\cdot 10^{-5}$ & 2$\cdot 10^{-2}$\\
45 & 4.0$\cdot 10^{-5}$ & -1.55 & 4$\cdot 10^{-6}$ & 2$\cdot 10^{-2}$\\
55 & 1.3$\cdot 10^{-4}$ & -1.98 & 2$\cdot 10^{-5}$ & 2$\cdot 10^{-2}$\\
60 & 6.4$\cdot 10^{-5}$ & -1.93 & 8$\cdot 10^{-6}$ & 2$\cdot 10^{-2}$\\ \hline
\end{tabular}
\label{fitpolD99}
\end{table}

Let us consider now the three patches from the U99 survey
at 1.4~GHz having low rotation measures, with centers and amplitudes given by:

\begin{eqnarray}
1 &:& {\rm l}=50^{\circ}, \
b=12^{\circ},\ \Delta {\rm l} =\ 10^{\circ},
\ \Delta b=15^{\circ},\nonumber\\
2 &:& {\rm l}=146.4^{\circ}, \
b=6.8^{\circ},\ \Delta {\rm l}=12^{\circ},
\ \Delta b=6.3^{\circ}.
\label{uyaregions}\\
3 &:& {\rm l}=196.8^{\circ}, \
b=11.2^{\circ},\ \Delta {\rm l} =\Delta b =\ 6.7^{\circ}\nonumber.
\end{eqnarray}

\begin{figure}
\resizebox{\hsize}{!}{\includegraphics{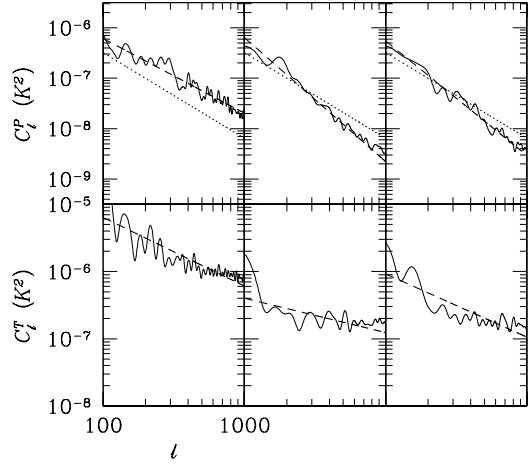}}
\vskip-40pt
\caption{Polarized (top) and total (bottom) intensity angular power spectra for
the three U99 areas at $1.4\,$GHz with low rotation measures specified by 
Eq.~(19): regions 1, 2, 3 from left to right. The dashed lines represent
power law fits. The dotted line shows the average power spectrum for the D97
survey, scaled in frequency according to Eq.~(18). For each panel, $\ell$
runs from 100 to 1000 on a logarithmic scale.}
\label{f8spectra}
\end{figure}

In Fig.~\ref{f8spectra} we plot the angular power spectra of these regions,
for polarized (top) and total (bottom) intensity.
As in the previous figures, the dashed lines show the power law fits
with the values of the parameters listed in Table~\ref{fituya}, and
the dotted lines represent Eq.~(\ref{fitduncan}).
Again there is a remarkable agreement with the power spectra derived
from the D97 survey, indicating that the polarized synchrotron emission
keeps essentially the same power spectrum (both in amplitude and in
slope) up to $b\simeq 10^{\circ}$.

\begin{table}
\caption{Power spectrum parameters of polarized (PI)
and total (I) intensity for the U99 survey at $1.4\,$GHz.}

\begin{tabular}{l c c c c} 
\hline
region & $\alpha\ ({\rm K}^{2})$ & $\beta$ &
$\Delta\alpha\ ({\rm K}^{2})$ & $\Delta\beta$\\ \hline
1 (PI) & 5.6$\cdot 10^{-4}$ & -1.48 & 8$\cdot 10^{-5}$ & 2$\cdot 10^{-2}$\\
2 (PI) & 5.3$\cdot 10^{-2}$ & -2.46 & 5$\cdot 10^{-3}$ & 2$\cdot 10^{-2}$\\
3 (PI) & 1.9$\cdot 10^{-2}$ & -2.27 & 2$\cdot 10^{-3}$ & 2$\cdot 10^{-2}$\\
1 (I) & 7.0$\cdot 10^{-4}$ & -1.02 & 1$\cdot 10^{-4}$ & 3$\cdot 10^{-2}$\\
2 (I) & 4.2$\cdot 10^{-6}$ & -0.51 & 7$\cdot 10^{-7}$ & 3$\cdot 10^{-2}$\\
3 (I) & 7.0$\cdot 10^{-5}$ & -0.94& 1$\cdot 10^{-5}$ & 3$\cdot 10^{-2}$\\
\hline
\end{tabular}
\label{fituya}
\end{table}

If we take into account also the power spectra obtained for the U99 regions
listed above and for the D99 regions at ${\rm l}\ge 50^{\circ}$, the average
power spectrum is described by:

\begin{eqnarray}
C^{P}_{\ell}&=& (1.2\pm 0.8)\cdot 10^{-9}\cdot
\left({\ell\over 450}\right)^{-1.8\pm 0.3}\cdot\nonumber\\
&\cdot&\left({\nu\over 2.4{\rm GHz}}\right)^{-5.8}\ {\rm K}^{2}\ ,
\label{fitall}
\end{eqnarray}
very close to that found from the D97 survey only [Eq.~(\ref{fitduncan})]

\subsection{High Galactic latitudes}

The accurate and high resolution D97, D99 and U99 polarization surveys
allow to understand the correlation properties of
polarized synchrotron emission at multipoles larger than $\sim 100$, owing to
the limited extent of the patches we can extract from the observed regions,
and cover regions at low or intermediate Galactic latitudes.

In order to extend the analysis of the polarized synchrotron fluctuation
power spectrum to superdegree angular scales and to high Galactic latitudes
we have exploited the BS76 polarization measurements.

We restrict  the present analysis to the 1.411~GHz channel, for a simpler
comparison with the high resolution surveys.
As we already mentioned in the Introduction,
at 1.411~GHz the BS76 measurements refer to 1726 positions over
approximately half of the sky at ${\rm l} \gsim 0^{\circ}$ with a beam FWHM of
about $0.6^{\circ}$. However, the sky is significantly undersampled and
the coverage is inhomogeneous.

We projected the original data into HEALPix maps
at different resolutions $\theta\simeq 3600/n_{\rm side}$
arcminutes, where $n_{\rm side}$ is the HEALPix resolution parameter
(G\'orski \etal~1998).
The whole map is essentially filled only at low resolutions,
$\theta \simeq 7.3^{\circ}$ or  $3.6^{\circ}$, corresponding to
$n_{\rm side} = 8$ or 16 respectively, that allow to estimate the angular power
spectrum only at $\ell\lsim 50$. On the other hand, we can take advantage of
some regions where the sky is better sampled and/or interpolate the existing
data to fill maps at higher resolutions,
$\theta \simeq 1.8^{\circ}$--$0.9^{\circ}$,
corresponding to $n_{\rm side} = 32$ or 64 respectively, to try an approximate
estimate of the power spectrum at higher multipoles, up to
$\ell\simeq 100$--$200$, and reach the multipole range where the power
spectrum estimation from high resolution surveys does not suffer of significant
boundary effects introduced by patch sizes. It is important in any case to fill
the map through interpolations to avoid the ``holes'' corresponding to
unobserved pixels. Such holes would introduce spurious (flat) power on the
pixel scale, being seen like ``negative sources'' in the power spectrum
estimation (see La~Porta \& Burigana 2000 for further details). We do not
expect crucial artifacts from this treatment, since the analysis of the
polarized synchrotron fluctuations on smaller angular scales presented above
indicates that the power significantly decreases toward high multipoles.
Of course, by comparing the power spectrum derived from the whole map
and from the regions where the sky is better sampled (and, consequently, the
possible interpolation effects less relevant) we can test the effect of these
approximations.

We identified three relatively well sampled regions in the maps derived from
the BS76 data:
1) a patch at low Galactic latitude ($110^{\circ} \le {\rm l} \le 160^{\circ}$,
$0^{\circ} \le b \le 20^{\circ}$), that can be used for a comparison with
the above results at higher resolutions and to extend the
analysis of the Galactic plane regions to low multipoles;
2) a relatively wide patch around the North Galactic Pole [($5^{\circ}
\le {\rm l} \le 80^{\circ}$, $b \ge 50^{\circ}$) together with
($0^{\circ} \le {\rm l} \le 5^{\circ}$, $b \ge 60^{\circ}$) and
($335^{\circ} \le {\rm l} \le 360^{\circ}$, $b \ge 60^{\circ}$) ];
3) a relatively small but better sampled region
($10^{\circ} \le {\rm l} \le 80^{\circ}$, $b \ge 70^{\circ}$),
included in the previous patch.
The analysis of the patches 2 and 3 allows to extend the polarized
synchrotron power spectrum estimation to high Galactic latitudes, where
the Galactic contamination is expected to be lower and,
correspondingly, the view of the CMB is cleaner (however,
the minimum of the Galactic emission is near ${\rm l} = 190^{\circ}$,
$b = 50^{\circ}$, see Berkhuijsen 1971).

\begin{figure*}
\resizebox{\hsize}{!}{\includegraphics{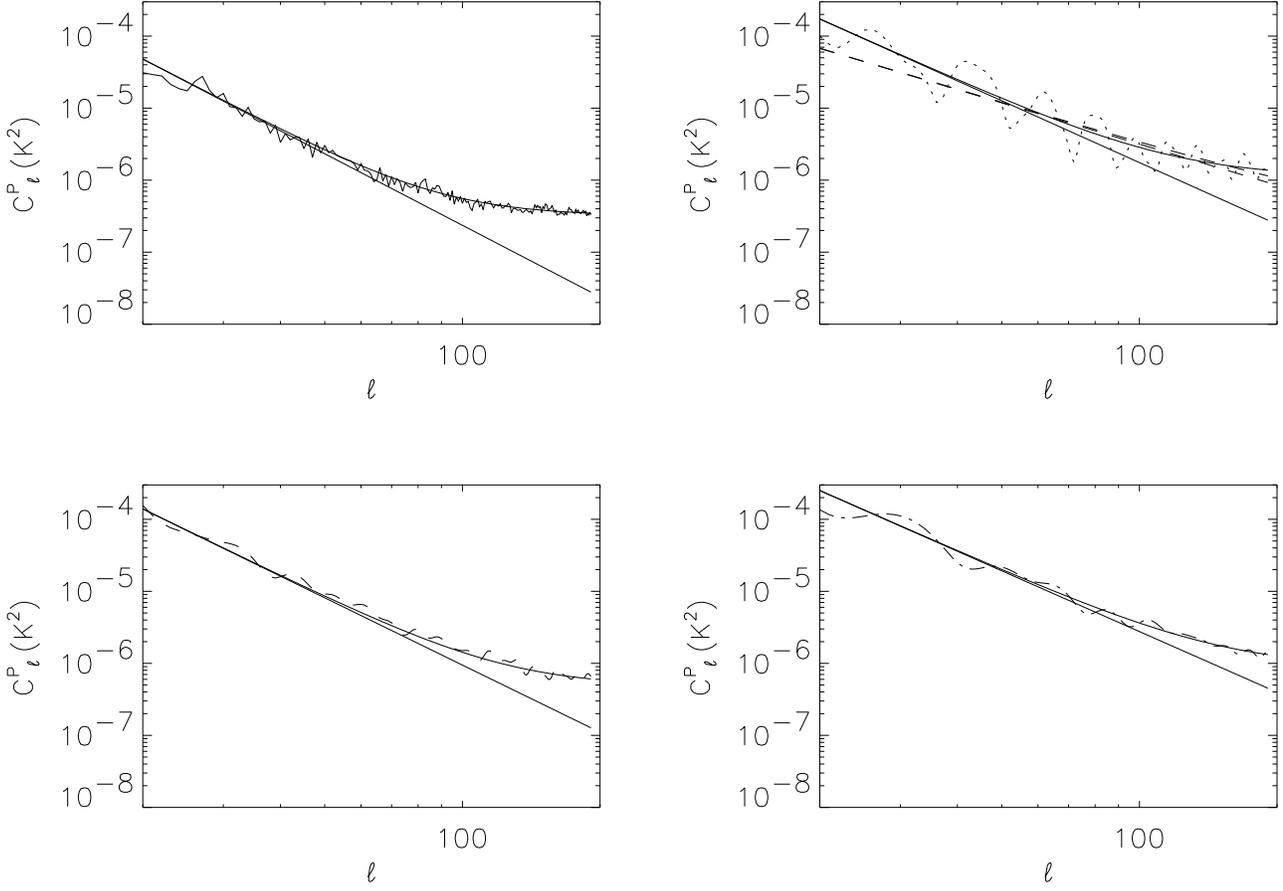}}
\caption{Fits of the angular power spectrum of polarized synchrotron emission
at 1.411~GHz derived by projecting into a map the Brouw \& Spoelstra (1976)
data for different sky regions.
Top left panel: whole map; top right panel: patch 1; bottom left panel:
patch 2; bottom right panel: patch 3.
Note the large oscillations found for the patch 1. The smooth solid
lines show the MIGRAD fits.
The smooth upper solid lines include also the noise contribution, whereas
the lower ones show the power law approximation
of the Galactic synchrotron component only, as derived from the fit.
SIMPLEX fits are very close to these for
patches 2 and 3 and are not shown; on the other hand, we show (dashed
line in the top right panel) the discrepant SIMPLEX fit for patch 1.
See the text for further details.}
\label{cf2}
\end{figure*}

The results, scaled to the case of a full sky coverage,
and the fits discussed below, 
are shown in Fig.~\ref{cf2}, for $30 \lsim\ell \lsim 200$.
At multipoles $\lsim 30$ boundary effects are
relevant, particularly for the patches.  At low $\ell$'s
the power spectrum decreases rather steeply with
increasing multipole number, then, at $\ell \gsim 100$, it flattens out.
Given the sensitivities quoted by BS76, it is likely that the
flattening is mostly due to instrumental noise.

To investigate the effect of the latter we have fitted the derived power
spectra as the sum of a power law plus white noise due to the instrument.
The best fit values of the parameters were computed using the MINUITS
software package of the CERN libraries
(available at the WEB site {\tt http://cern.web.cern.ch/CERN/}).
We neglected the effect of beam smoothing, which becomes important
at multipoles higher than those at which the noise power starts to be
dominant.

In order to test the stability of the results, we have carried out
the calculations using two different routines:
MIGRAD and SIMPLEX. The results are very close to each other except for
patch 1, as discussed below. For the other three cases (whole map and
patches 2 and 3) we present only the fits produced by
MIGRAD, while both fits are given for patch 1 (see also Fig.~\ref{cf2}).

MIGRAD yields:

\begin{equation}
C_\ell^{P}
\simeq (9.4 \cdot 10^{-1} \cdot\ell^{-3.3} +3.2\cdot 10^{-7})\ {\rm K^{2}}
\label{clbsmap}
\end{equation}

for the whole map,

\begin{equation}
C_\ell^{P}
\simeq (1.5 \cdot\ell^{-3.1} +4.8\cdot 10^{-7})\ {\rm K^{2}}
\label{clbsp2}
\end{equation}

for patch 2,

\begin{equation}
C_\ell^{P}
\simeq (1.1 \cdot \ell^{-2.8} +8.7\cdot 10^{-7}) \ {\rm K^{2}}
\label{clbsp3}
\end{equation}

for the patch 3, and
\begin{equation}
C_\ell^{P}
\simeq (8.8 \cdot 10^{-1} \cdot\ell^{-2.85} +1.1\cdot 10^{-6})\ {\rm K^{2}}
\label{clbsp1}
\end{equation}

for the patch 1. For the latter patch the SIMPLEX fit is

\begin{equation}
C_\ell^{P}
\simeq (2.0 \cdot 10^{-2} \cdot\ell^{-1.9}
+2.2\cdot 10^{-7})\ {\rm K^{2}}\, .
\label{clbsp1bis}
\end{equation}

As illustrated by Fig.~\ref{cf2}, the power spectrum of polarized
emission for patch 1 shows substantial oscillations, which are responsible for
the instability of the power law fit derived with the two routines. We believe
the MIGRAD result to be more reliable for at least two reasons: first, it is
consistent with that derived from the data of U99 for the area common to
both surveys, for the overlapping range of scales ($\ell \sim 200$), while
the SIMPLEX fit yields a signal a factor of several too high; second,
the estimated noise level implied by the SIMPLEX fit seems to be too low,
being substantially lower than found for the other two patches and even
smaller than found for the whole map (in all these cases SIMPLEX and
MIGRAD results are in close agreement), 
whereas the estimated noise level implied by the MIGRAD fit
is in agreement also with that derived by inverting a patch of noise
simulated on the basis of the sensitivities quoted by BS76
in the sky directions of the considered region.

If we can ignore the SIMPLEX result for patch 1, as due to a numerical
instability in the presence of large oscillations of the power spectrum,
we may conclude that the power spectra for the three regions agree 
with each other to within a factor of about 2. Remarkably, no 
significant differences are found between the region on the Galactic 
plane and the North polar region.
The power law terms in Eqs.(\ref{clbsp2})--(\ref{clbsp1}) are all 
consistent with

\begin{eqnarray}
C^{P}_\ell &=& (4\pm 3) \cdot 10^{-7} \cdot
\left({\ell\over 50}\right)^{-2.9\pm 0.2}\cdot\nonumber\\
&\cdot&\left({\nu\over 2.4{\rm GHz}}\right)^{-5.8}\ {\rm K^{2}}\, ,
\label{clbsfit}
\end{eqnarray}

where, to ease the comparison with Eq.~(\ref{fitall}), the amplitude was
scaled to $2.4\,$GHz using the typical synchrotron spectrum.

The power spectrum found for the whole map exhibits a slope
quite close to that of the three patches considered and an amplitude
about 3--6 times smaller, mostly due to the lower polarization degree
in the regions far from the patches considered here.

In spite of their poor sky sampling and of all the approximations
introduced to treat them, the BS76 data indicate an amplitude of
the polarization power spectrum in the region of overlap
($\ell \simeq 100$) quite close to that
derived from more recent surveys, although the slope is steeper.
This suggests that, at least on superdegree angular scales, the
fluctuations of the diffuse Galactic polarized synchrotron emission
are almost independent of Galactic latitude.

\begin{figure}
\resizebox{\hsize}{!}{\includegraphics{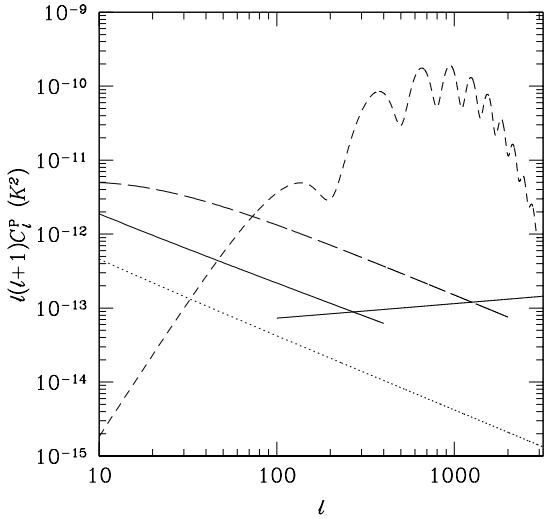}}
\vskip-15pt
\caption{Comparison of the power spectrum of the CMB polarized component
predicted by a standard CDM model (dashed line) with the power spectrum
of the polarized Galactic synchrotron emission at 100 GHz, as
yielded by Eqs.~(21) and (27) for high and low values of $\ell$, respectively
(solid lines). Also shown are simple minded estimates made assuming that
the synchrotron emission has everywhere the maximum theoretical polarization
degree ($\sim 75\%$): the long-dashed and the dotted lines correspond
to the average power spectrum of the total power synchrotron emission
at high Galactic latitudes ($|b|>30^\circ$) derived by one of us (C. Burigana;
see also Tegmark \& Efstathiou 1996) from the map of Haslam et al. (1982),
and to the median power spectrum estimated by Bouchet \& Gispert 
(1999; see also Bouchet \etal~1999), respectively.}
\label{cmb}
\end{figure}

\section{Discussion and conclusions}

\label{conclusion}

We have analyzed the available surveys of diffuse Galactic polarized
emission at GHz frequencies. The most recent ones cover regions
at low and medium Galactic latitudes with angular resolution $\simeq 10'$
or better (Duncan \etal~1997, 1999, Uyaniker \etal~1999).
Observations at high Galactic latitudes have a resolution of $0^\circ.6$
(Brouw \& Spoelstra 1976).

A particularly bewildering issue is Faraday depolarization, which may
be substantial at the frequencies of the data analyzed here (a few GHz) but
will be negligible at 100 GHz. Space varying Faraday depolarization effects may
substantially affect the power spectrum of polarized synchrotron emission
in several ways: the amplitude is decreased and structure may be created
on a variety of scales. If so, the validity of extrapolations of the present
results to MAP's and {\sc Planck}'s frequencies may be largely spoiled.
However, our careful analysis of depolarization effects has led to the
identification of regions where rotation measures of pulsars and
extragalactic sources, the polarization degree and, in some cases,
data on the distribution of polarization vectors and on the Galactic
magnetic field, consistently indicate that Faraday 
depolarization must be small.

The low polarization degree of radio emission close to the Galactic plane 
is interpreted as due to large
contributions to the observed intensity from unpolarized sources, primarily
strong H{\sc ii} regions. Since such sources are concentrated on the
Galactic plane, estimates of the power spectrum of total intensity at
low Galactic latitudes are not representative of the spatial distribution of
Galactic emission far from the plane.

Since we need to extrapolate the results by a large factor in frequency
(from a few GHz up to 100 GHz) a second important issue is the spectral index
of synchrotron emission to be adopted. The average spectral index
of the antenna temperature between 408 MHz and 7.5 GHz is about 2.8 (Platania
\etal~1998); it is expected to steepen to $\simeq 3$ above $\simeq 10\,$GHz
as a consequence of the steepening of the energy spectrum of cosmic rays
above $\simeq 15\,$GeV, due to energy losses (Banday \& Wolfendale 1990, 1991).
Indications of a steepening were indeed found by Platania \etal~(1998)
already at 7.5 GHz and from their comparison of the 408 MHz map by Haslam
\etal~ (1982) with a preliminary map at 19 GHz (Cottingham 1987;
Boughn \etal~1990). A further steepening is expected at still higher
frequencies. Thus, our choice of an average spectral index of 2.9 up to
100 GHz is rather on the conservative side: it probably leads to an
overestimate of the amplitude of the synchrotron power spectrum at this
frequency. On the other hand, it must be kept in mind that there are evidences
(Reich \& Reich 1988; Platania \etal~1998) of flatter spectral indices at
higher Galactic latitudes, where CMB maps are expected to be the cleanest.
Also, synchrotron spectral indices show substantial spatial variations,
which will yield a high frequency power spectrum significantly different
from the one estimated from the data analyzed here.

Yet another issue is the extrapolation of our results to different regions of
the Galaxy, particularly to higher Galactic latitudes.
In general, we expect that the power spectrum of total and polarized emission
varies across the sky. For example, it is reasonable to expect less small
scale structure in the general anticenter region because emission cells
have generally larger angular sizes, being, on average,
relatively closer.

It is likely that the mean amplitude of the power spectrum of
Galactic emission decreases with increasing Galactic latitude as a
consequence of the decreased emission. Also, as pointed out by Davies \&
Wilkinson (1999),
the magnetic field pattern may be more ordered at high Galactic latitudes,
resulting in less small scale structure. Moreover, narrow depolarizing
structures, that may be present in regions where depolarization is
generally small, may again increase the amplitude of the the polarization
fluctuations on small scales observed at relatively low frequencies.
Since polarization surveys cover only a limited fraction of the sky and,
moreover,
observations at high Galactic latitudes allow to estimate the
polarization power spectrum only on super-degree scales, all these issues
remain, to a large extent, open.

However, our results do suggest that the dependence on Galactic 
latitude of the power spectrum of polarized synchrotron emission 
is weak. Polarization fluctuations are found to remain rather
low even  at relatively low Galactic latitudes. This is very encouraging since
the highest sensitivity polarization maps that will be produced by the
{\sc Planck} satellite will cover regions around the Ecliptic poles, which are
at moderate Galactic latitudes ($|b|\simeq 27^\circ$).

In spite of all these uncertainties, our analysis
suggests that the Galactic polarization fluctuations
are unlikely to hinder MAP's and {\sc Planck}'s measurements of
CMB polarization for $\ell\ge 50$. This is illustrated by Fig.~\ref{cmb} where
the expected level of CMB polarization fluctuations predicted by
a standard CDM model is compared with our estimates of the Galactic
contamination at $100\,$GHz [Eqs.~(\ref{fitall}) and (\ref{clbsfit})].
The same figure also shows the average power spectrum of the total power
synchrotron emission at high Galactic latitudes ($|b|>30^\circ$)
derived by one of us (C. Burigana; see also Tegmark \& Efstathiou 1996)
from the map of Haslam \etal~(1982), scaled
to the maximum theoretical polarization degree ($\sim 75\%$),
corresponding to a uniform magnetic field (long dashed line).
This polarization
degree is a conservative upper limit, since turbulent components of the
magnetic field decrease the polarization degree. Also, there are evidences
of significantly lower synchotron emissions over relatively large regions of
the sky. In Fig.~\ref{cmb} this is illustrated by the dotted line, showing
the median synchotron power spectrum estimated by Bouchet \& Gispert 
(1999); see also Bouchet \etal~1999) from the 1420 MHz survey of Reich 
\& Reich (1988), extrapolated to higher frequencies with a spectral index 
of 2.9; again a polarization degree of $75\%$ has been adopted.
The CMB polarized component is usually decomposed in suitable eigenfunctions
keeping memory of different kinds of cosmological perturbations,
generally known as E and B modes (see Hu \etal~1998 and references therein).
We find that the Galactic synchrotron emission discussed here contributes
almost equally to the two modes, as expected since Galactic and CMB
signals are completely uncorrelated.

Finally, the distributions of both total and polarized Galactic emissions was
shown to be non-Gaussian at a high significance level. This may be a
problem for methods of component separation using Wiener filtering, which
assumes Gaussian distributions. On the other hand, it allows the application
of Independent Component Analysis techniques (Baccigalupi \etal~2000)
which just rely on the assumption that all but at most one of the components
to be separated have a non-Gaussian distribution.

\begin{acknowledgements}
Thanks are due to Drs. A.R. Duncan and B. Uyaniker who have generously made
available their maps through the Web and also provided additional information
on their results. We are grateful to Dr. T.A.T. Spoelstra for his kind
clarifications. Thanks are also due to the referee, Dr. E.M. Berkhuijsen,
for her very careful reading of the manuscript and for very useful
comments that led to substantial improvements of the paper.
This work was supported in part by ASI and MURST.
\end{acknowledgements}

\end{document}